**Oxysterols in vascular cells and role in atherosclerosis**

Celine Luquain-Costaz[1,2] and Isabelle Delton[1,2]

[1] CNRS 5007, LAGEPP, Université of Lyon, Université Claude Bernard Lyon 1, 69100 Villeurbanne, France.
[2] Department of Biosciences, INSA Lyon, 69100 Villeurbanne, France.

Corresponding author : Pr. Isabelle Delton, isabelle.delton@insa-lyon.fr

**ABSTRACT**

Atherosclerosis is a major cardiovascular complication of diseases associated with elevated oxidative stress such as type 2 diabetes and metabolic syndrome. In these situations, low density lipoproteins (LDL) undergo oxidation. Oxidized LDL display proatherogenic activities through multiple and complex mechanisms which lead to dysfunctions of vascular cells (endothelial cells, smooth muscle cells and macrophages). Oxidized LDL are enriched in oxidized products of cholesterol called oxysterols formed either by autoxidation, enzymatically, or by both mechanisms. Several oxysterols have been shown to accumulate in atheroma plaques and to play a key role in atherogenesis. Depending on the type of oxysterols, various biological effects are exerted on vascular cells to regulate the formation of macrophage foam cells, endothelial integrity, adhesion and transmigration of monocytes, plaque progression and instability. Most of these effects are linked to the ability of oxysterols to induce cellular oxidative stress and cytotoxicity mainly through apoptosis and proinflammatory mediators. Like for excess cholesterol, high density lipoproteins (HDL) can exert antiatherogenic activity stimulating the efflux of oxysterols that have accumulated in foamy macrophages.

**Abbreviations**

CT : cholestane-3β-5α-6β-triol ; 4β-OHC : 4β-hydroxycholesterol ; 7KC : 7-ketocholesterol ; 7α-OHC : 7α-hydroxycholesterol ; 7β-OHC : 7β-hydroxycholesterol ; 5,6α-EC : 5α,6α-epoxycholesterol ; 5,6β-EC : 5β,6β-epoxycholesterol ; 20S-OHC : 20(S)-hydroxycholesterol ; 22R-OHC : 22(R)-hydroxycholesterol ; 22S-OHC : 22(S)-hydroxycholesterol ; 24S-OHC : 24(S)-hydroxycholesterol ; 25-OHC : 25-hydroxycholesterol ; CH25H : cholesterol 25-hydroxylase ; 27-OHC : 27-hydroxycholesterol ; IL : interleukin ; MMP : matrix metalloproteinase ; oxLDL : oxidized low density lipoprotein ; EC : endothelial cells ; ER : endoplasmic reticulum ; SMC : smooth muscle cells ; TNFα : tumor necrosis factor ; HDL : high density lipoprotein

1. Introduction

Atherosclerosis is a chief cause of morbidity and mortality in Western countries. It is characterized by the formation in the intima arteries of atherosclerotic plaques consisting of the accumulation of lipids, complex carbohydrates, blood cells and products, fibrous tissue and calcium deposits. Since the regions with low fluid shear stress - the sites of arterial branching or curvature - are much more susceptible, the abdominal aorta, coronary arteries and carotid bifurcations are the predilection vessels for the formation of atherosclerotic plaques. The formation of atherosclerotic plaques also called atherogenesis is a long term and evolutionary processus. Initial lesions (fatty streaks) can be observed very early in life and stay asymptomatic for a long time. Advanced atherosclerotic lesions develop with the accumulation of lipids, cell debris and fibrous tissue in a long term processus (20 to 40 years) The thickening of the plaques cause cardio-vascular diseases (CVD) such as coronary heart disease, myocardial infarction, and stroke when disrupting. The risk factors for atherosclerosis are multiple including non-modifiable factors such as age, sex and genetic background, and modifiable factors such as smoking, lack of exercise, high-fat diet. It is well acknowledged that atherosclerosis is a major cardiovascular complication of diseases associated with chronic inflammatory status, increased oxidative stress and disorder of lipid metabolism, such as type 2 diabetes, obesity and metabolic syndrome.

Arteries are constituted of three morphologically distinct layers : the outerlayer adventitia, the media, and the innermost layer intima where the atherosclerotic plaques are formed. Atherogenesis begins with the infiltration and subendothelial accumulation of low density lipoproteins (LDL). In physiopathological situations, LDL become oxidized (oxLDL) closely related to oxidative stress prevailing in the arterial wall that causes oxidation of the protein and lipid moieties of LDL through both enzymatic or non-enzymatic (ROS-induced) pathways. LDL oxidation can be catalyzed by metal cations (cupper, iron) and several enzyme systems, including 12/15-lipoxygenase strongly expressed strongly in atherosclerotic plaques, myeloperoxidase, NOS (nitric oxide synthase), xanthine oxidase, and NADPH oxidase. The oxidative products are various, including oxidized derivatives of fatty acid (e.g. malondialdehyde (MDA), 13-HPODE, 13-HODE), and lipids (e.g. lysophosphatidylcholine, oxysterols) as well as protein carbonyls. The early oxidation of LDL can only produce the minimally oxidized LDL (m-oxLDL) because of the presence of antioxidants, such as vitamin E, A and β carotene, and the efficient anti-oxidant enzymes (e.g. superoxide dismutase). In the narrow sub-endothelial space, LDL undergo further oxidation to form the highly oxidized LDL (h-oxLDL) involving the ROS produced by macrophages and ECs, and also several enzymes form these cells. OxLDL display atherogenic activities through multiple and complex mechanisms. Globally, oxLDL altered homeostasis of vascular cells resulting in loss of endothelial cell integrity, SMC migration and proliferation and foam cell formation through different signals linked to proinflammatory and proapoptotic effects (Nègre-Salvayre et al. 2017; Negre-Salvayre et al. 2020).

The LDL receptor (LDL-R) is located on the plasma membrane and internalizes LDL after binding to apoB-100. At a minor stage of oxidation (m-oxLDL), the apoB is simply no longer recognized by the LDL-R. At a major stage of oxidation (h-oxLDL), the modified apo-B allows the recognition of multiple scavenger receptors (SR) including SR-A1 (scavenger receptor class A1), SR-A2, CD36, SR-B1 (scavenger receptor class B1) and LOX1 (lectin-like oxidized LDL receptor 1) (Zingg et al. 2021). In contrast with LDL-R, even though the uptake of modified LDL induces the elevated cholesterol content in macrophage, the expression of SR is not regulated by the intracellular cholesterol quantity, which results in high uptake of oxLDL by macrophages and cholesterol (as its ester form) accumulation in macrophages. The cholesterol-engorged macrophages are called foam cells. The accumulation of foam cells in artery walls, which is easily recognized by light microscopy, is the sign of early atherosclerotic lesions.

Among oxidized lipids in oxLDL, cholesterol oxidation products of cholesterol also called oxysterols have recently gained special attention with respect to their role in atherogenesis. This review is an overview on the oxysterols that accumulate in atheroma plaques and that exert biological activities on vascular cells including endothelial cells, smooth muscle cells and macrophages, thereby contributing to atherogenesis and ultimate plaque rupture.

## 2. Oxysterols in atherosclerotic plaque and in serum from patients with CVD

As mentioned above, oxLDL are enriched in oxysterols, some of which being involved in the ability of oxLDL to induce cellular oxidative stress and cytotoxicity, mainly through apoptosis. Oxysterols are also associated with the regulation of lipid metabolism, inflammation and are considered as factors contributing to clinical complications of atherosclerosis (see paragraph 3).

Oxysterols are produced from cholesterol oxidation through enzymatic pathways with oxidative position mainly on the side chain of cholesterol generating hydroxycholesterols (OHC) including 20S-OHC, 22R-OHC, 22S-OHC, 24S-OHC, 27-OHC, 25-OHC, and 7α,25-dihydroxycholesterol. Non-enzymatic pathways cholesterol autoxidation mainly on the cholesterol ring led to 7-ketocholesterol (7-KC), 7β-OHC, 7α-OHC, 4β-OHC, 5α,6α-epoxycholesterol (5,6α-EC), 5β,6β-epoxycholesterol (5,6β-EC) and cholestan-3β,5α,6β-triol (CT). 7α-OHC and 25-OHC can be formed by both pathways.

### 2.1. Levels of oxysterols in atheroma plaque and plasma of atherosclerotic subjects

The first evidence for the presence of oxysterols in atherosclerotic plaques dated from mid 60s (Brooks et al. 1966; Fumagalli et al. 1971). Increased amounts 27-OHC and 7-KC were reported in these studies. Since then, many studies have confirmed the association between the presence of oxysterol in atheromatous lesions and atherosclerotic plaque formation, progression and stability. Overall, 27-OHC is the major oxysterol recovered in advanced atherosclerotic plaques, followed by 7-KC, 7β-OHC and 7α-OHC. These oxysterols account for 75-85% of the total oxysterols detected in atherosclerotic plaques from various sites, the others being 25-OHC, 24-OHC and 5,6-EC (Vaya 2013).

Elevated levels of 27-OHC were found in human carotid atherosclerotic plaques compared to non-atherosclerotic human vessels in correlation to high expression of 27-hydroxylase (the enzyme converting cholesterol to 27-OHC) in macrophage-rich core regions of complicated lesions (Crisby et al. 1997). Since 27-OHC can be eliminated from macrophages, it was proposed that 27-OHC formation could be a defense mechanism against deleterious cellular accumulation of cholesterol.

Analysis of oxidized lipids in human aortic advanced atherosclerotic plaques revealed elevated amounts of 27-OHC and 7β-OHC compared to normal aorta (Carpenter et al. 1993). This group also highlighted the presence of these two oxysterols at different stages of atherosclerotic lesions (fatty streaks, intermediate and advanced lesions) from aorta and common carotid artery, compared to human normal artery. 27-OHC was significantly more abundant in advanced lesions than in intermediate lesions or fatty streaks (Carpenter et al. 1995). 27-OHC and 7β-OHC were also detected in all the samples of human atheromatous lipid core and fibrous cap of individual advanced atherosclerotic plaques (Garcia-Cruset et al. 1999). It is also noteworthy that pharmacological lowering of 27-OHC was associated with coronary plaque regression (Nakano et al. 2022).

7-oxygenated sterol (7β-OHC, 7α-OHC and 7KC) have also been widely detected in atherosclerotic lesions; 7-KC is the second most abundant oxysterol and the major 7-oxygenated sterol found in atheromatous plaques(Brown et al. 1997; Garcia-Cruset et al. 2001; Ravi et al. 2021), although it was not detected in a recent study (Pinto, 2022). The presence of 7β-HOC in atherosclerotic lesions together with 27-OHC has been reported in early studies (Carpenter et al. 1993; Garcia-Cruset et al. 1999). Importantly, the *in situ* formation of ROS-derived 7-oxygenated sterol in human carotid plate was confirmed - rather than auto-oxydation during sample processing - as well as that of ROS-derived EC (Helmschrodt et al. 2013).

Other oxysterols such as 24-OHC, 25-OHC, 5β,6β-EC, 5α,6α-EC were found in human fatty streaks and advanced atherosclerotic lesions (Garcia-Cruset et al. 2001; Helmschrodt et al. 2013). Increased levels of 24-OHC, 25-OHC, 27-OHC, 7α-OHC and 7β-OHC but not 7-KC were measured in symptomatic subjects with carotid atherosclerotic plaques compared to asymptomatic subjects(Pinto et al. 2022). Arterial intima accumulation of 27-OHC and 24S-OHC is associated with severe peripheral artery disease (Virginio et al. 2015). The accumulation of 25-OHC was reported in coronary atherosclerotic lesions (Canfrán-Duque et al. 2023) but not in arterial tissue from subjects with severe peripheral artery disease (Virginio et al. 2015).

In addition to high detection in atheromatous plaques, elevated plasma concentrations of several oxysterols including 25-OHC, 27-OHC and 7β-OHC were shown to correlate with symptoms of coronary and peripheral artery disease and atherogenic risk profile (Ziedén et al. 1999; Yasunobu et al. 2001; Rimner et al. 2005; Virginio et al. 2015). Increased levels of 7β-OHC in plasma was proposed to be a biomarker for high risk of developing cardiovascular disease and coronary atherosclerotic plaques (Khatib and Vaya 2014). Elevated plasma 7-KC levels have also been associated with higher risk of cardiovascular disease events in the general population and in patients with coronary artery disease (Hitsumoto et al. 2009; Song et al. 2017; Wang et al. 2017).

Atherosclerosis is a major complication of diseases associated with high oxidative stress such as diabetes, metabolic syndrome and dyslipidemia. The level of total oxysterols was markedly increased in the serum of diabetic subjects compared to healthy controls, mainly due to high amounts of 7-KC, 7α-OHC, 7β-OHC and α-EC (Khatib and Vaya 2014). Elevated plasmatic oxysterol levels especially 7-KC and CT were correlated to hyperglycemia and glycation index as well as number of coronary risk factors, particularly in type 2 diabetic patients (Samadi et al. 2019; Samadi et al. 2020; Ahmed et al. 2022).

**2.2. In vitro and in vivo formation of oxysterols**

In vitro oxidation of LDL using different acellular oxidizing conditions (copper, peroxynitrite, 2,2'-azobis(2-amidinopropane)dihydrochloride-AAPH, hypochlorous acid) has confirmed the formation of oxysterols in the particle through auto-oxidation process, including 7-KC, 7α-OHC, 7β-OHC that were reproducibly detected. The formation of 5α,6α- and 5β,6β-EC, as well as 24-OHC, 25-OHC, or 27-OHC has also been reported but to a lesser extent and not in all studies (Matsunaga et al. 2009; Vaya et al. 2011; Arnal-Levron et al. 2013; Orsó et al. 2015; Chen et al. 2015b). The level of oxysterols in LDL varies according to the oxidizing conditions and is the highest in copper highly oxidized LDL (Orso, 2015).

Although less explored, cell-mediated generation of oxysterols was also reported. Macrophages, as well as EC and SMC, were shown to mediate radical dependent LDL oxidation leading to the formation of lipid hydroperoxydes and consequent increase of 7-OHC and 7-KC (Müller et al. 1998). It is commonly assumed that accumulation of oxysterols in atheroma macrophages results from the massive uptake of oxLDL that itself contain oxysterols (Brown et al. 1996; Brown et al. 1997). Accordingly, the levels of 7-KC, 7α-OHC, 27-OHC and cholest-4-en-3-one were found to increase in oxLDL-loaded mouse macrophages proportionally to the degree of LDL oxidation (Paul et al. 2019).

The intracellular formation of oxysterols in cultured macrophages is very low and usually under the detection limit but some species (*i.e.* 7-KC, 25-OHC, 24-OHC, 24,25-EC) were highly increased upon excessive cholesterol loading or exposure to the endotoxine Kdo2-Lipid A (Fu et al. 2001; Dennis et al. 2010; Maurya et al. 2013). Intracellular formation of oxysterols is promoted in macrophages after uptake of modified LDL that undergo oxidation inside lysosomes leading to the formation of 7-KC and 7β-OHC (Wen and Leake 2007; Yoshida and Kisugi 2010). Our studies confirmed that cellular activity significantly contributes to the accumulation of oxysterols in macrophage cell lines (Arnal-Levron et al. 2013; Chen et al. 2015b). Upon exposure to copper-oxidized LDL, both cellular cholesterol and LDL-derived cholesterol were oxidized in murine RAW and human THP1 macrophages resulting in a huge increase of oxysterol production. Major oxysterols originated from non-enzymatic pathway (7-KC and 7α/β-OHC), the enzymatically formed 25-OHC and 27-OHC being recovered in much lower proportions. We also demonstrated that the oxidation of LDL-derived cholesterol occurred mainly in the late endosomal compartment, while oxidation of cellular cholesterol likely occurs at the plasma membrane site (Chen et al. 2015b). The intracellular oxysterol production in oxLDL-loaded RAW macrophages was regulated by the specific endosomal phospholipid bis(monoacylglycero)phosphate, also supporting LDL-cholesterol oxidation in this compartment (Arnal-Levron et al. 2013).

3. **Biological effects of oxysterols in atherogenesis and plaque progression.**

Several oxysterols have been involved in various aspects of atherogenesis, such as formation of macrophage foam cells, endothelial dysfunction, adhesion and transmigration of monocytes, plaque progression and instability (Poli et al. 2009). Depending on the type of oxysterols, various biological activities are exerted on vascular cells including EC, SMC and macrophages. Although a subset of oxysterols are likely to exert antiatherogenic effects by regulating cholesterol homeostasis, an overlapping but distinct set of oxysterols display proatherogenic activity by inducing cytotoxicity primarily through apoptosis and stimulating inflammatory pathways.

**3.1 Oxysterols and foam cell formation**

Cholesterol homeostasis in macrophage results from the finely regulated balance between cholesterol acquisition including *de novo* synthesis of cholesterol dependent on HMGCoA reductase (HMGCoAR) activity and uptake of non oxLDL by LDL-R, and efflux of excess cholesterol by extracellular acceptors HDL and apoA1. Both mechanisms of cholesterol acquisition are regulated by a complex of three proteins located in the endoplasmic reticulum (ER): SREBP (sterol regulatory element binding protein), SCAP (SREBP cleavage activating protein) and Insig (insulin induced gene). This complex regulates the expression of LDL-R receptor and HMGCoAR inversely correlated to ER cholesterol content (Sato 2010; Savla et al. 2022). The removal of excess intracellular cholesterol is regulated by the activity of nuclear receptor liver X receptors (LXR) by enhancing the expression of ATP-binding cassette (ABC) transporters ABCA1 and ABCG1 (Matsuo 2022). Atherosclerosis is characterized by excessive accumulation of cholesterol within subendothelial macrophages. To sum up, cholesterol accumulates consequently to the massive unregulated uptake of oxLDL by SR and to the defects in ABC dependent efflux of cholesterol (Li et al. 2021). In early lesions, excess cholesterol is stored in the form of cholesterol esters produced by acyl-coA:cholesterol ester transferase (ACAT) activity within lipid droplets giving macrophages their foamy appearance. In advanced lesions, free cholesterol accumulates in ER, leading to ER stress that triggers apoptosis of foamy macrophages .

Side chain oxysterols, in particular 25-OHC, suppress the expression of SREBP2 target genes presumably in an Insig-dependent manner, which negatively regulates cholesterol biosynthesis and uptake (Sato 2010). As natural ligands for LXRα and LXRβ (Janowski et al. 1999), side chains oxysterols including 24,25-EC, 22(R)-OHC, 24(S)-OHC, 25-OHC and 27-OHC exert antiatherogenic activity by stimulating the expression of ABCA1 and ABCG1 in macrophages (Töröcsik et al. 2009; Olkkonen 2012; Saito et al. 2023). However, some oxysterols can inversely promote foam cell phenotype. A mixture of 7β-OHC and 7-KC stimulates lipid droplet formation and upregulates class A scavenger receptor that facilitates the ingestion of oxLDL (Yuan et al. 2016; Ward et al. 2017; Saha et al. 2020)).

### 3.2 Oxysterol and apoptosis/autophagy/oxyautophagy

Apoptosis is a natural, programmed cell death process that occurs in multicellular organisms. Two main pathways trigger apoptosis: the intrinsic pathway (or mitochondrial pathway) and the extrinsic pathway (or death receptor pathway). Both pathways involve a series of complex signaling events that ultimately lead to the activation of caspases and the subsequent destruction of the cell. The intrinsic pathway is activated by intracellular stresses, such as DNA damage, oxidative stress, or nutrient deprivation. It is associated with outer mitochondrial membrane permeabilization and release of cytochrome c that induces the assembly of a caspase-activation complex. The extrinsic pathway is activated by the binding of extracellular ligands, such as Fas ligand or tumor necrosis factor-alpha (TNFα) to death receptors on the cell surface which initiates the caspase cascade. Caspase 3 is considered to be a key mediator of apoptosis; it cleaves a number of key cellular proteins, including cytoskeletal proteins, nuclear proteins, and enzymes, leading to DNA fragmentation, and membrane dismantling (D'Arcy 2019).

The ability of oxysterols to induce apoptosis in vascular cells has been well described. Among the oxysterols found in atheroma plaques, 7-OHC and 7-KC are commonly the most cytotoxic. These oxysterols induce apoptosis in SMC (Hughes et al. 1994; Miyashita et al. 1997; Pedruzzi et al. 2004)), EC (Luchetti et al. 2017) and macrophages (Li et al. 2012; Ward et al. 2017; Ravi et al. 2021) contributing to the cause of cell death in core regions of atherosclerotic plaques. Compared to 7-oxysterols, side chain oxysterols exert no or lower apoptotic effect. 27-OHC exerts dual effects in terms of cytotoxicity towards macrophages, acting as a protector at a low concentration while triggering apoptosis at high concentrations (Riendeau and Garenc 2009). Low micromolar concentration of 27-OHC evokes survival signals in U937 human promonocytic cell line through the activation of ERK and AKT and inhibition of the proapoptotic protein Bad in response to initial ROS formation (Vurusaner et al. 2016; Vurusaner et al. 2018). Studies using 25-hydroxylase deficient macrophages suggest that 25-OHC increases susceptibility to LPS induced apoptosis in macrophages through increased caspase 3 activation and reduced efferocytosis capacity (Canfrán-Duque et al. 2023).

7-Oxysterols stimulate both intrinsic mitochondrial and extrinsic death receptor pathways of apoptosis. In macrophages, apoptosis mediated by 7-oxysterols is associated with caspase-3 activation, increased permeability of mitochondrial membrane, release of cytochrome c and endonuclease G (Prunet et al. 2005; Palozza et al. 2010; Li et al. 2012). 7-KC and 7β-OHC can also induce ROS production and decrease cellular antioxidants, therefore inducing a mitochondrial oxidative stress in the sub-endothelial space (Tabas et al. 2015; Ravi et al. 2021). Apoptosis mediated by 7-oxysterols also involves the regulation of the expression of Bcl-2 family proteins that exert pro- or anti-apoptotic activity. In macrophages, 7-KC induces proapoptotic pathways associated with the proapoptotic proteins Bax and Bim (Berthier et al. 2005; Palozza et al. 2010; Li et al. 2012) and inhibits the anti-apoptotic protein AKT (Vejux and Lizard 2009; Palozza et al. 2010). However, 7-KC was also reported to trigger a survival response through the activation of PYK2/MEK1/2/ERK1/2 pathway allowing BAD phosphorylation (Berthier et al. 2005). 7-KC and 7β-OHC induced apoptosis is also associated with the induction and nuclear translocation of the tumor suppressor p53 (Li et al. 2012) which has been shown to be highly expressed and to promote apoptosis in human atherosclerotic plaque (Yuan et al. 2010).

Endoplasmic reticulum (ER) stress is a cellular response to an accumulation of misfolded or unfolded proteins within the ER. ER stress triggers a signaling pathway called the unfolded protein response (UPR) that aims to restore ER homeostasis. ER markers include phosphorylated eIF2α and IRE1α and expression of the proteins GRP78 and CHOP. Prolonged activation of UPR leads to cell dysfunction and death, contributing to the development of various diseases, including neurodegenerative disorders, diabetes, cancer, and cardiovascular diseases.. The links between oxLDL and ER stress as well as the involvement of ER stress in atherosclerosis initiation and progression have been well documented, but the role of oxysterols is not well established (Sanda et al. 2017; Luchetti et al. 2017). In SMC and EC, 7-KC induced ER stress as characterized by increased phosphorylation of IRE1α and expression of CHOP and GRP78 (Pedruzzi et al. 2004; Sanson et al. 2009). 7-KC and 7-OHC can induce macrophage apoptosis through moderating ER stress-specific signaling involving CHOP induction (Son et al. 2012; Park et al. 2016).

Autophagy is a process by which cells degrade and recycle damaged or unwanted cellular components to promote cell survival. Autophagy has been recently revealed as a crucial regulator in the formation of early and advanced atherosclerotic plaques (Li et al. 2022). This phenomenon is traditionally regarded as beneficial in atherosclerosis as it prevents EC apoptosis and senescence, regulates the proliferation of SMC cells and inhibits foam cell formation and lipid-laden macrophage apoptosis. Few studies have examined the role of oxysterols in autophagy of vascular cells. 7-KC was reported to induce autophagy in human SMC promoting survival and stabilizing atherosclerotic plaque (He et al. 2013; Zhang et al. 2020). On the other hand, 7-KC-induced autophagy may exert deleterious effects as it promotes vascular calcification through autophagy-lysosomal pathway (Sudo et al. 2015).

Last decade, cell death induced by oxysterols has been defined as a complex mode of cell death involving oxidative stress, apoptosis and autophagy, defined as oxiapoptophagy Oxiapoptophagy is associated with organelle dysfunction and in particular with mitochondrial and peroxisomal alterations involved in the induction of cell death and in the rupture of redox balance. Oxidative stress can induce both apoptosis and autophagy, and these processes can interact with each other. For example, under certain conditions, autophagy can promote apoptosis by degrading anti-apoptotic proteins, while apoptosis can inhibit autophagy by cleaving essential autophagy proteins. Thus, the interplay between oxidative stress, apoptosis, and autophagy is complex (Nury et al. 2014; Nury et al. 2021). With respect to oxysterols 7-KC, 7β-OHC and 24S-OHC were reported to be strong inducers of oxiapoptophagy in different cell types including U937 monocytic cells (Nury et al. 2021; de Freitas et al. 2021). Other oxysterols have been shown to induce oxiapoptophagy such as 24(S)OHC in oligodendrocytes (Nury et al. 2015), 25-OHC in fibroblast cells (You et al. 2021), and 7α,25-dihydroxycholesterol in osteoblasts (Seo et al. 2023). However, the link between oxysterols and oxiapoptophagy in atherosclerosis is not documented.

**3.3. Oxysterols and Inflammation**

Oxysterols are considered as potent inducers of inflammation as they can alter endothelial monolayer integrity, recruit the immune cells, stimulate the expression of various inflammatory molecules, regulate macrophage polarization into M1/M2 phenotypes and promote the rupture of fibrous caps (Testa et al. 2018).

It is well known that endothelial dysfunction occurs in the early stage of atherosclerosis and some oxysterols have been involved in this process. 25-OHC is able to inhibit EC proliferation, migration, and tube formation and to impair endothelium-dependent vasodilation through inhibition of NO production (Ou et al. 2016). It also reduced the expression of tight junction proteins (Niedzielski et al. 2021). 25-OHC, as well as 7-KC, was also shown to reduce endothelial monolayer impedance and adhesion (Chalubinski et al. 2013).

Integrins are components of cell-matrix adhesions which regulate monocyte adhesion to endothelial cells and their migration to the site of inflammation. The expression of β1-integrin was shown to increase in U937 monocytes exposed to a oxysterol mixture of pathophysiologic relevance through activation of the G-protein/c-Src/PLC/PKC/ERK signaling pathway. Among oxysterols present at concentrations close to those found in vascular lesions, 7α-OHC and CT were the most potent compounds and 7β-OHC, 5α,6α-EC, 7-KC the less potent ones (Gargiulo et al. 2012). The Oxysterol mixture was previously shown to stimulate the expression and synthesis of MCP-1 (monocyte chemotactic protein-1), another monocyte chemoattractant also referred to as CCL2, in U937 macrophages through the activation of ERK1/2 (extracellular signal-regulated kinase 1/2) pathway and nuclear binding of NF-κB (nuclear factor κB) (Leonarduzzi et al. 2005). It was also reported that 25-OHC is a ligand of α5β1 and αvβ3 integrins to activate integrin-focal adhesion kinase (FAK) signaling (Pokharel et al. 2019).

ICAM-1 (intercellular adhesion molecule-1), VCAM-1 (vascular cell adhesion molecule-1) and E-seletin are adhesion molecules involved in the interaction between leukocyte and EC and subsequent transmigration of monocytes through the endothelial monolayer. Oxysterols especially those oxidized at C7 (7α/β-OHC and 7-KC), increased the levels of ICAM-1, VCAM-1 and E-selectin expression in human vascular cells through mechanisms involving p38MAPK pathway or ROS production (Lemaire et al. 1998; Shimozawa et al. 2004; Tani et al. 2018). 25-OHC is also able to induce ICAM-1 synthesis in human EC associated with disruption of endothelial integrity, both effects being counteracted by statin (Niedzielski et al. 2021).

Oxysterols were also reported as modulators of proinflammatory cytokines such as interleukins IL1,IL6, IL8. In macrophages, 7β-OHC, 7-KC and 25-OHC regulate IL8 production involving MEK/ERK pathway and AP-1 mediated process (Erridge et al. 2007; Lemaire-Ewing et al. 2009). IL8 induction and secretion in macrophages was also reported after treatment with 7α-OHC through mechanisms dependent of C5a receptor and PI3K and MEK pathways (Cho, 2017). 7-KC also stimulates IL12 and IL1α in macrophages (Saha et al. 2020) and 27-OHC stimulates the secretion of IL8 and IL1 in human monocytes through activation of TLR4/NFKB pathway (Gargiulo et al. 2012; Kim et al. 2013). In EC, 7-KC stimulates the expression and secretion of IL-8 associated with ROS production and PI3K/Akt signaling pathways (Chang et al. 2016). 25-OHC also induces proinflammatory cytokines in EC including IL1β, IL-18, IL-23 while anti-inflammatory cytokines IL-10 and IL-37 were repressed (Woźniak et al. 2023). TNFα which is mainly expressed in macrophages is a key regulator of the cytokine cascade and displays proatherogenic activity by promoting the formation of foam cells. The production of TNFα in macrophages is stimulated by 27-OHC and 7α-OHC but not 7-KC in macrophages (Kim et al. 2013; Gargiulo et al. 2015). 25-OHC promotes the production of proinflammatory cytokines including TNF and IL6 through activation of αvβ3 integrin signaling (Pokharel et al. 2019).

<span style="color:red">As natural ligands for LXRα and LXRβ , side chains oxysterols including 24,25-EC, 22(R)-OHC, 24(S)-OHC, 25-OHC and 27-OHC exert antiatherogenic activity by check effect on cytokines?   (Töröcsik et al. 2009; Olkkonen 2012; Saito et al. 2023).</span>

Many studies have also highlighted the key role of oxysterols in promoting atherosclerotic plaque instability and rupture (Gargiulo et al. 2018). Matrix metalloproteinases (MMPs) belong to the zinc-metalloproteinases family which are involved in the degradation of extracellular matrix, therefore playing a key role in the rupture of fibrous caps and the formation of advanced atherosclerotic lesions. MMP are primarily produced by activated macrophages but also by vascular SMC and EC. Recently, the role of pro-protein convertase subtilisin/kexin protease (PCSK) 6 in plaque instability and rupture has been suggested related to its ability to stimulate the activity of MMP and its elevated expression in symptomatic carotid plaque. Among oxysterols from a mixture representative of those present in advanced human carotid plaques, 27-OHC, 7α-OHC and to a lesser extent 25-OHC were shown to be the most potent inducers of MMP-9 expression in human monocytes through the activation of TLR4/NFkB pathway (Gargiulo et al. 2012; Gargiulo et al. 2015). It was further demonstrated that the mixture of oxysterols induced MMP-9 activity through PCSK6 activation in monocytes (Testa et al. 2021). In SMC, 7-KC and cholesterol-5α, 6α-epoxide induce the expression of MMP-2 and MMP-9 through the EGFR/PI3K/Akt signaling pathways, which was associated with SMC migration and proliferation (Liao et al. 2010). In atherosclerotic mice, oxysterol-rich diet, containing mainly 7C-derivatives and EC, induces plaque instability and rupture related to increased MCP1 expression and MMP activity (Sato et al. 2012).

Macrophages exist as two main subsets, the classically activated macrophages-pro-inflammatory M1 phenotype and the alternatively activated macrophages-anti-inflammatory M2 phenotype. M1 macrophages are primarily found in rupture-prone atherosclerotic plaques, while alternatively activated macrophages accumulate in stable plaque. 7-KC increased the production of the pro-inflammatory cytokines TNF-α and IL-6 in M1 macrophages (Buttari et al. 2013). In addition, 7-KC is able to redirect the polarization of M2 macrophages to an M1-like subset by changing the profile of surface markers and cytokines towards an anti-inflammatory phenotype, by decreasing endocyte clearance capacity and by increasing the secretion of MMP9 secretion (Buttari et al. 2013; Buttari et al. 2014; Saha et al. 2020). By contrast, a mixture of oxysterols as well of 27-OHC which is the major oxysterol present in the mixture was reported to favor plaque stabilization by driving M2 polarization of human macrophages (Marengo et al. 2016). After entering the cells through CD receptors, oxysterols trigger LXR activation which leads to IL10 secretion and MIF release and contributes to atherosclerotic plaques stabilization.

Other inflammatory mediators such as prostaglandines (PG) whose production is regulated by the enzyme COX2 contribute to plaque development. COX2 expression and synthesis as well as PGE2 production are increased by CT in EC (Liao et al. 2009). 27-OHC promotes up-regulation of COX-2 and PGE synhase thereby inducing PGE2 synthesis in human monocytes. This regulation is associated with enhanced production of pro-inflammatory cytokines including IL8, IL1β and TNFα and of MMP9, which leads to plaque instability (Gargiulo et al. 2018).

### 4. Mechanisms of oxysterols efflux

It is well acknowledged that plasma high-density lipoprotein (HDL) levels are inversely related to the risk of atherosclerotic cardiovascular disease, related to multiple anti-atherosclerotic functions exerted by HDL such as antioxidative capacity, anti-inflammatory activity, cytoprotective activity and protection on endothelium‑dependent vasorelaxation. The best known atheroprotective activity of HDL is its ability to export cholesterol from foamy macrophages and artery wall. Transport proteins ABCA1 and ABCG1 play a central role in cholesterol efflux: ABCA1 regulates the efflux of cholesterol from the macrophage to apoA1/pre-β HDL while ABCG1 mediates efflux to mature HDL.

We and others have shown that HDL can also promote the efflux of oxysterols that accumulated in macrophages after exposure to oxLDL (Terasaka et al. 2007; Xu et al. 2009; Iborra et al. 2011; Chen et al. 2015a; Paul et al. 2019). Overall, 7-KC and 7α/β-OHC were the most efficiently released and 7-KC-induced macrophage apoptosis was reduced, which confers protective effect to HDL against pro-atherogenic oxysterols. In contrast to HDL, ABCA1-dependent apoA1 exhibits weak ability to export oxysterols including 7-KC from macrophages (Kritharides et al. 1995; Gelissen et al. 1999; Terasaka et al. 2007; Xu et al. 2009; Chen et al. 2015a). Another study reported opposite results showing that apoA1 efficiently export 7-KC, 7α-OHC, 5,6-α-EC, cholest4-en-3-one and 27-OHC from oxLDL-loaded macrophages while HDL only exerted a trend toward oxysterol efflux (Paul et al. 2019).

In some pathological situations such as type 2 diabetes and obesity with elevated risk of atherosclerosis, HDL undergo modifications including oxidation, glycation and glycoxidation, which impair ability to efflux cholesterol (Denimal, 2023). In the study by Chen et al (Chen et al. 2018), we showed that oxidized and glycoxidized HDL as well as HDL isolated from diabetic subjects, have decreased ability to efflux oxysterols from oxidized LDL-laden macrophages compared to HDL from healthy subjects. Efflux of 7-KC was specifically decreased compared to that of cholesterol. This defect of HDL towards oxysterol efflux may potentiate the deleterious effects of oxysterols and especially 7-KC that accumulate in atheroma macrophages. Oxysterols were detected at very low levels in HDL of healthy subjects, mainly represented by 7-KC, and in lower proportions 7α/β-OHC, 25-OHC and 27-OHC. Similar or even lower amounts were found in HDL from diabetic subjects, which could reflect their strong antioxidant capacities that would protect cholesterol from oxidation, or their weaker capacity to mobilize cellular oxysterols in particular 7-KC. Under high oxidative conditions used to generate oxidized and glycoxidized HDL, a considerable increase of 7C-derived oxysterols, in particular 7-KC, is observed (Chen et al. 2018). It was not evaluated whether the high amounts of oxysterols in these modified HDLs are responsible for their lower ability to remove oxysterols, as it has been proposed towards cholesterol efflux (Gesquière et al. 1997).

5. **Conclusions**

The involvement of oxysterols in atherosclerotic plaque formation and instability is now well established. Among oxysterols most abundant in atheroma plaque, 7-derivative oxysterols in particular 7-KC, 5,6-EC, and CT exert proatherogenic effects through induction of proapoptotic and proinflammatory mediators. Side chain oxysterols in particular 25-OHC and 27-OHC display dual effects as they also exert protective effects by preventing foam cell formation, or plaque instability. It is now well demonstrated that oxidative stress is determinant for the formation of oxysterols as well as signaling pathways evoked by deleterious oxysterols. The cellular targets of oxysterols including membrane receptors, signaling pathways and transcription factors are also well documented. Therapeutic strategies to counteract deleterious effects of oxysterols in atherosclerosis and other diseases such as cancer and neurodegenerative diseases have started to be evaluated either based , pharmacological inhibition of oxysterol activated signaling pathways (Lee et al. 2015; Park et al. 2016; Saha et al. 2020), oxysterol derivatives (de Medina et al. 2021) or antioxidants naturally present in nutrition oils (Nury et al. 2021; Rezig et al. 2022). Pharmacology of oxysterols is a promising route for the development of new drugs against current high incidence diseases and thus deserves further investigations.

**Conflict of interest**

The authors declare no conflict of interest

**Acknowledgments**

This work was supported by funding from INSA Lyon

**Figure legends**

Figure 1 : Regulation of cell death by oxysterols.7-KC, 7-ketocholesterol; 7β-OHC, 7β-hydroxycholesterol, ER Stress, endoplasmic reticulum stress.

Figure 2 : Regulation of inflammation by oxysterols. ICAM, intercellular adhesion molecule-1 ; VCAM, vascular cell adhesion molecule, TNFα, tumor necrosis factor α; MMPs, Matrix metalloproteinases; SMC, smooth muscle cells